\documentclass{iopart}
\usepackage{iopams}
\usepackage{color}
\usepackage{graphicx}

\newcommand{\bB}{\mathbf{B}}

\newcommand{\bj}{\mathbf{j}}
\newcommand{\bk}{\boldsymbol{\kappa}}
\newcommand{\bu}{\mathbf{u}}

\newcommand{\epol}{\mathbf{e}_\theta}
\newcommand{\etor}{\mathbf{e}_\xi}
\newcommand{\fsa}[1]{\left\langle{#1}\right\rangle}
\newlength{\figwidth}
\newcommand{\eqref}[1]{(\ref{#1})}
\newcommand{\rev}[1]{{#1}}

\begin{document}

\title{Dynamics of zonal flow-like structures in the edge of the TJ-II stellarator}

\author{J.A.~Alonso, J.L.~Velasco, J.~Ar\'evalo, C.~Hidalgo, M.A.~Pedrosa, B.Ph.~Van~Milligen and D.~Carralero}
\address{Euratom-CIEMAT Assoc., Laboratorio Nacional de Fusi\'on, Av Complutense 22, 28040 Madrid, Spain}
\author{C.~Silva}
\address{Euratom-IST Assoc., Instituto de Plasmas e Fus\~ao Nuclear, Instituto Superior T\'ecnico, Universidade T\'ecnica Lisboa, Lisboa, Portugal}

\ead{ja.alonso@ciemat.es}

\begin{abstract}
The dynamics of fluctuating electric field structures in the edge of the TJ-II stellarator, that display zonal flow-like traits, is studied. These structures have been shown to be global and affect particle transport dynamically \cite{AlonsoNF2012}. In this article we discuss possible drive (Reynolds stress) and damping (Neoclassical viscosity, geodesic transfer) mechanisms for the associated $E\times B$ velocity. We show that: (a) while the observed turbulence-driven forces can provide the necessary perpendicular acceleration, a causal relation could not be firmly established, possibly because of the locality of the Reynolds stress measurements, (b) the calculated neoclassical viscosity and damping times are comparable to the observed zonal flow relaxation times, and (c) although an accompanying density modulation is observed to be associated to the zonal flow, it is not consistent with the excitation of pressure side-bands, like those present in geodesic acoustic oscillations, caused by the compression of the $E\times B$ flow field.

\end{abstract}

\pacs{52.25.Os,52.30.-q,52.55.Hc}

\maketitle

\section{Introduction}
Mass flows are an active research topic in magnetic confinement fusion due to their importance for plasma stability and confinement.
In recent years, both experimental and theoretical efforts have been made to improve understanding of the momentum transport mechanisms that determine the observed plasma rotation profiles in tokamaks (see, e.g., \cite{TalaPPCF2007}). In stellarators without a direction of symmetry in the magnetic field strength, the neoclassical non-ambipolar fluxes are formally dominant in determining the equilibrium radial electric field and plasma rotation. Mass flows in such a system experience a magnetic viscous damping in all directions, which tends to keep their amplitude low \cite{HelanderPRL2008}.

Fluctuating, flux surface-constant zonal flows have a specific interest, for they are thought to be instrumental in turbulent transport control and to be a natural state of the drift-wave type turbulence. The question: to what extent different magnetic configurations can host zonal flow structures? has gathered some attention \cite{SugamaPRL2005, MynickPoP2007, HelanderPPCF2011}. For non-symmetric systems, Sugama and Watanabe \cite{SugamaPRL2005} found the suggestive result that neoclassical optimization could lead to reduced ZF damping and a turbulent transport optimization as a `welcome side-effect'. Previously, this had been found experimentally in LHD's inward-shifted configuration \cite{MotojimaNF2003}. 

Experimental evidence of long range correlations (LRCs) in distant electric potential signals was first found in the CHS stellarator \cite{FujisawaPRL2004} with a double Heavy Ion Beam Probe system and later in the TJ-II stellarator with a double Langmuir probe system \cite{PedrosaPRL2008}. Shortly aferwards other devices reported similar findings (see \cite{XuNF2011} and references therein).

In this article we continue the anaysis of the spatio-temporal characteristics of zonal flow-like floating potential structures initiated in Ref.\cite{AlonsoNF2012}. In that reference, a method to extract the ZF component from a set of LRCed signals was introduced and tested. The reconstructed structure showed actual traits of a zonal flow as a collective, fluctuating and transport-regulating structure (see next section). Our concern in this work is with the dynamics of these structures: their drive and damping mechanisms. We investigate the involvement of the Reynolds stress in the generation of the zonal-flow structures and the neoclassical collisional damping and geodesic transfer as possible damping mechanisms.

\section{Experimental set-up description}
Experiments where carried out in the 4-period, flexible, low shear stellarator TJ-II using electron cyclotron heated plasmas. The typical magnetic field strength is $\sim 1$ T and the rotational transform $\frac{\iota}{2\pi}(a) = 1.65$ at the last closed flux surface (LCFS) (typical minor radius $a$ is 20 cm), dropping slowly to a central value of $\frac{\iota}{2\pi}(0) = 1.55$. Electron temperatures range from $1-0.8$ keV at the centre to an edge value of $100 - 50$ eV.%
\begin{figure}
\includegraphics[width=\figwidth]{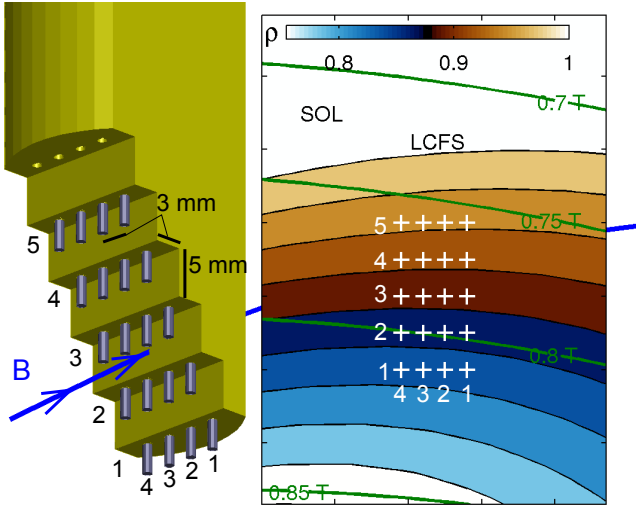}%
\caption{\label{fig:probe}Pin arrangement in the 2D probe and orientation with respect to the magnetic field structure of TJ-II.}
\end{figure}
The floating potential measurements presented in this work were obtained with a 2D array of $5\times4$ tungsten pins, with radial and poloidal separations of $6$ and $3$ mm, respectively (see figure \ref{fig:probe}). Simultaneous measurements are taken with another pin in a different sector of the machine, having a long parallel connection length with the 2D probe ($\gtrsim 100$ m).The sampling rate of the floating potential signals is 2 MHz.

\section{Analysis of the dynamics of the zonal-flow like events}
In previous work \cite{AlonsoNF2012} we introduced a method to extract the spatial and temporal characteristics of the zonal flow-like component of the long-distance correlated floating potential signals. This extraction is based on the Singular Value Decomposition. Here we give a brief explanation of the results of that analysis and refer the interested reader to the original reference for details.

A large matrix $M$ if formed defining its entries as
\[
 M_{ij} = \phi_i(t_j)~,
\] 
where $\phi_i(t_j)$ is the $i$-th floating potential evaluated at the $j$-th time sample. Here $t_j = (j-1)\Delta t$ and $(\Delta t)^{-1} = 2\times10^6$~Hz. The size of $M$ is thus $m\times n$, where $m$ is the number of floating potential signals ($20+1$ in our case) and $n$ is the number of samples in the time window under analysis ($6\times 10^4$ in our case).

The SVD of $M$ can be written as the sum
\begin{equation}\label{eq:svd2}
	M  = \sum_\alpha \sigma_\alpha \mathbf{u}^\alpha(\mathbf{v}^\alpha)^\dag~.
\end{equation}
where $\{\mathbf{u}^{\alpha}\}_{\alpha=1\ldots m}$ and $\{\mathbf{v}^{\alpha}\}_{\alpha=1\ldots n}$ are orthonormal vector sets and are called \emph{topos} and \emph{chronos} respectively. The spatial information is coded in the topos set whereas the temporal evolution is contained in the chronos set. The singular values $\sigma_\alpha$ are non-negative and are sorted in decreasing order, i.e. $(\sigma_1 \ge \sigma_2\ge\ldots\ge 0)$. Their squared values are the contributions of each of the modes $\alpha$ to the total signal energy, i.e. $E =\sum_{ij}\left|\phi_i(t_j)\right|^2 \equiv \sum_{\alpha=1}^{\mathrm{rank}(M)}\sigma_\alpha^2$

In \cite{AlonsoNF2012} we tentatively identified the zonal flow component as
\begin{equation}\label{eq:defzf}
\Phi_{ZF} =  \sigma_1\mathbf{u}^1(\mathbf{v}^1)^\dag~,
\end{equation}
This identification was motivated by the empirical observations that this perfectly correlated spatio-temporal structure was: (a) Low-$k_\theta$, (b) broadband spectrum but dominated by low ($\lesssim 10$~kHz) frequencies \rev{(no clear peak indicative of GAMs was observable)} (c) explained most of the observed long-range correlation. Consistently, we then showed that (d) it was collective, i.e. fluctuations occurred in the 2D probe and the distant probe simultaneously and (e) the zonal flow amplitude defined as $A_{ZF}(t_j) = (v^1)^*(t_j)$ dynamically modulated the outward particle transport as shown by all H-alpha monitors around the device. \rev{We refer the reader to the cited reference for a more detailed explanation of these points.}

\subsection{Conditionally averaged $E_r$ excursion and dynamical parameters}
\begin{figure}
\includegraphics{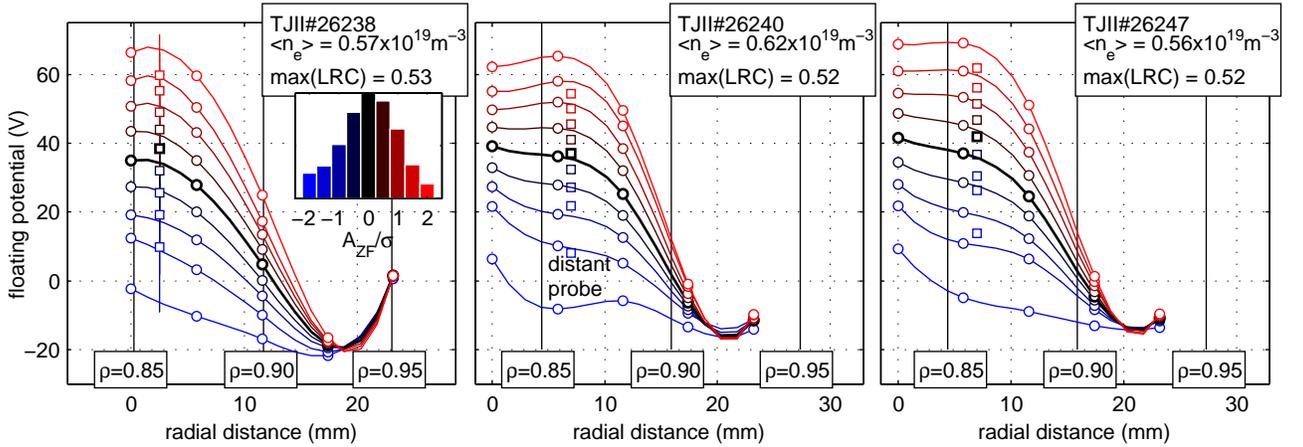}
\caption{Long range correlated, zonal flow-like floating potential modulations for three TJ-II shots. Each profile is conditionally average to $A_{ZF}$ values in the different regions of its pdf as shown in the inset. The conditionally averaged distant probe potential is shown with squares. The bars in the left figure correspond to 1 standard deviation.}
\label{fig:profiles}
\end{figure}

\begin{figure}
\includegraphics{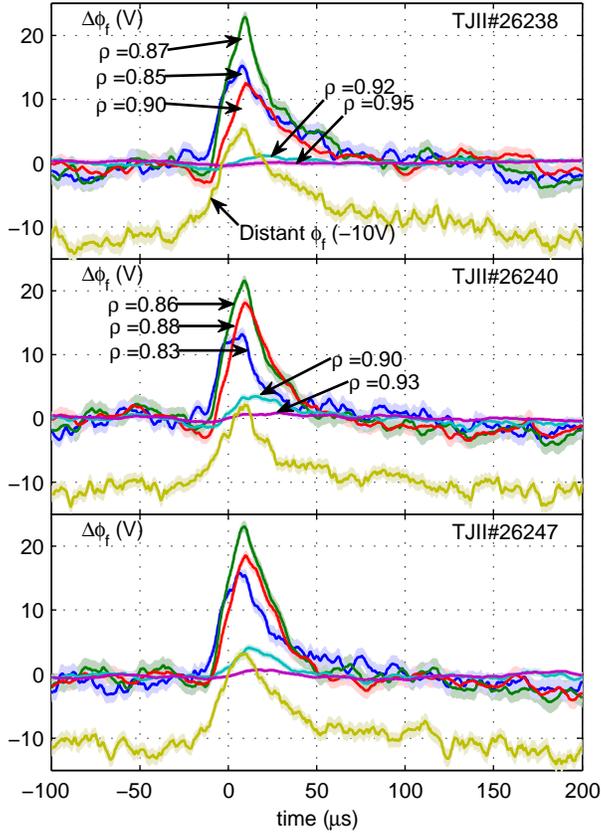}
\caption{Typical evolution of the floating potential excursions shown in Fig.\ref{fig:profiles}. Each trace correspond to the poloidal average of the floating potential signals at each radius. This spatial average is then de-offset and conditionally averaged to $A_{ZF}/\sigma> 1$). Similar evolution is seen in the distant probe signals, showing the collective character of the excursions.\label{fig:waveforms}}
\end{figure}

In this section we make use of the extraction technique presented above to study the temporal and spatial characteristics of the ZF structure. Regarding the spatial structure we are interested to know what is the radial profile of the potential modulation and also to verify our definition (Eq.\ref{eq:defzf}) showing that the modulations are flux-surface collective. This is done in figures \ref{fig:profiles} and \ref{fig:waveforms}.  
 
Figure \ref{fig:profiles} shows the spatial shape of the floating potential profile modulations for three different discharges. Each radial point in the profiles is obtained as the poloidal average of the floating potentials conditionally averaged to $A_{ZF}$ values in different regions of its probability density function (PDF, shown in the inset). The different colours code different deviations from the mean, i.e. most likely, profile shown in black. 
The modulations in the distant probe are plotted with square symbols located around the radial position of the peak LRC. In these shots the distant probe was held fixed. It displays modulations of the same sign though somewhat lower amplitude as those in the 2D probe.

In figure \ref{fig:waveforms} we show how the floating potential excursions typically (conditionally averaged to $A_{ZF}/\sigma > 1$) proceed in time. Again, the collective nature of the excursions is observed in the similar waveforms of the 2D and distant probe evolutions. A temporal asymmetry is noticeable in these structures. In the next sections we obtain the dynamical parameters associated with the time evolution of $E\times B$ velocity and compare the result with turbulent Reynolds stress and Neoclassical viscosity as possible drive and damping mechanisms.

\subsection{Relation to turbulent Reynolds stress forces}
\begin{figure}
\includegraphics{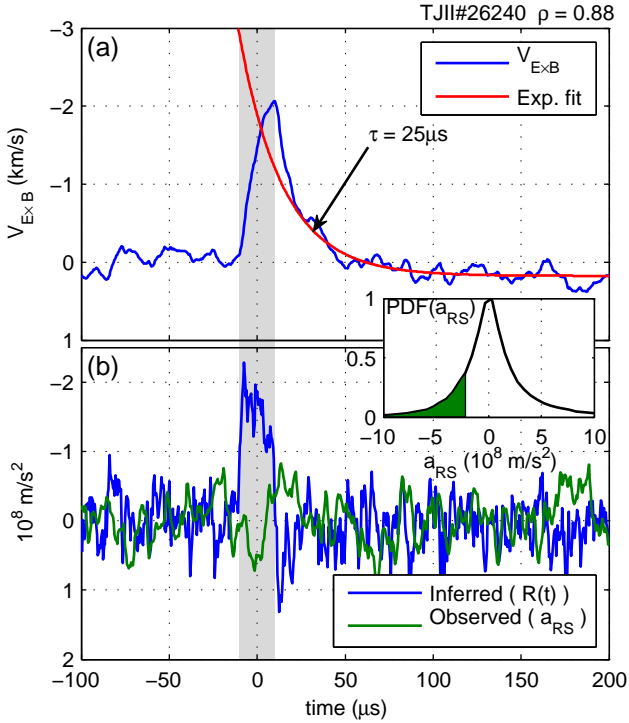}
\caption{(a)Typical $E\times B$ velocity excursion at $\rho = 0.88$ for shot TJII\#26240, showing a decay time of $\sim 25 \mu s$. (b) Comparison of the inferred forcing (Eq.\ref{eq:model}) with the measured Reynolds stress acceleration. The probability density function of the RS acceleration shows the events capable of providing excursions like the one in (a) or larger.\label{fig:dynamics}}
\end{figure}

A typical $E\times B$ velocity excursion is shown in figure \ref{fig:dynamics}.a. This is a derived quantity from the floating potential evolution shown in Fig.\ref{fig:waveforms} for the shot TJII\#26240 by subtracting the traces at $\rho=0.86$ and $\rho=0.90$ to get an estimate of the radial electric field at $\rho= 0.88$. We model the evolution as the competition of a drive term $R(t)$ and a magnetic viscous damping term, i.e.
\begin{equation}\label{eq:model}
 \frac{d}{dt}V_E = R(t) - \nu V_E~.
\end{equation} 
We can get an estimate of the damping time $\nu^{-1}$ by fitting an exponential to the $V_E$ decay in figure \ref{fig:dynamics}.a. \rev{The clear peak in the typical floating potential evolution (Fig.\ref{fig:waveforms}) is indicative of a sudden change in the dynamics (e.g. the forcing) of the structure}. We therefore assume that flow is freely decaying and the forcing $R$ vanishes on average (recall this waveform is conditionally averaged) for $t\gtrsim 30\mu s$.  The so obtained decay times is $\nu^{-1} = 25 \mu s$, similar to the one obtained in \cite{AlonsoNF2012} and in biasing relaxation experiments \cite{PedrosaPPCF2007}. This value is then used to integrate Eq.\ref{eq:model} to infer the acceleration $R(t)$ assuming the viscosity is constant in time (fig.\ref{fig:dynamics}.b). A time-dependent viscosity has been derived in \cite{ShaingPoP2005} which shows that the effective collision frequency involved in the viscosity is augmented by the addition of the growth rate of the ZF. Using this simplifying hypothesis we ignore this correction, and note that in general this leads to an underestimation of the inferred acceleration.

A natural question is what physical mechanism provides the required force or acceleration. The pin distribution of 2D probe allows to compute the turbulent flux of poloidal (or more precisely perpendicular) momentum $\Gamma_{v_\theta} = \tilde{v}_{E,r}\tilde{v}_{E,\theta}$ (i.e. the perpendicular Reynolds Stress (RS)) at the location of the six central pins. We can then estimate the RS acceleration, given by the divergence of the flux, as
\begin{equation}
a_{RS} = - \frac{d}{dr}\fsa{\tilde{v}_{E,r}\tilde{v}_{E,\theta}} 
\end{equation}
where, the radial derivative is approximated by finite differences and the poloidal average $\fsa{\cdot}$ is the mean of the poloidally separated $\Gamma_{v_\theta}$ measurements at each radial position. Figure \ref{fig:dynamics}.b shows the conditional average of $a_{RS}(\rho = 0.88)$ over the same time windows used to get the $V_E$ evolution shown in \ref{fig:dynamics}.a. There is no observable causal relation between the \emph{locally} measured $a_{RS}$ and the $V_E$ excursions. The inset in the figure shows the probability density function of $a_{RS}$, which nevertheless shows that the inferred acceleration fall within the support of the PDF. In other words: we observe RS boosts capable of providing the required acceleration, were they global, i.e. representative of the flux surface averaged RS, but observe no temporal correlation between those boosts and the $V_E$ excursions. The analysis of the two other shots casts similar conclusions.

\rev{We emphasise that this conclusion is subject to several experimental approximations in the calculation of the Reynold stress acceleration. First, we are implicitly taking a poloidaly and toroidaly localised measurement to be representative of the flux surface average. Second, we are approximating the fluctuations in the plasma potential by those of the probe floating potential, thus neglecting electron temperature fluctuations.}

\subsection{Comparison of relaxation time with the Neoclassical collisional damping rate}
We begin this section by introducing a basic model of flow and viscosity. We follow Ref.\cite{WobigPPCF1995} in the notation and choice of Hamada coordinates $(V, \theta, \xi)$. In this coordinates the magnetic field is written as
\begin{equation}
\bB = 2\pi\chi'(V)\epol + 2\pi\psi'(V)\etor~,
\label{eq:bfield}
\end{equation}
where $\chi$ and $\psi$ are the poloidal and toroidal magnetic fluxes respectively and prime ($'$) denotes derivation with respect to their argument (the volume enclosed by the flux surface $V$). The $s$-species flows are assumed to be dynamically incompressible and tangent to flux surfaces. They are compactly written as
\begin{equation}
\bu_s = E_s(\psi)\epol + \Lambda_s(\psi)\bB~,
\label{eq:flow}
\end{equation}
in terms of two flux constants $E$ and $\Lambda$. The former can be written in terms of the $s$-pressure $p_s$ and electric potential $\phi$ as $E_s(\psi) = 2\pi\left(\frac{p'(\psi)}{Z_sne} + \phi '(\psi)\right)$. With this definition, and the property $\epol\times\bB = \frac{\nabla\psi}{2\pi}$ (valid for any magnetic coordinate system) the perpendicular $	E\times B$ plus diamagnetic flow is recovered form $\bB\times$~(Eq.\ref{eq:flow}). The parallel part of $\epol$ is the Pfirsch-Schlueter flow, as the Hamada poloidal covariant base vector satisfies $\nabla\cdot\epol = 0$ and \rev{$\fsa{\epol\cdot\bB} =  (\mu_0/2\pi)I_{T}$} ( $= 0$ for a zero toroidal current stellarator). The $\Lambda_s$ term is then the so-called `bootstrap' flow $\Lambda_s = \fsa{\bu_s\cdot\bB}/\fsa{B^2}$. 

In our extraction of the zonal flow dynamical parameters, namely acceleration and relaxation time, we assumed a freely decaying flow and fitted an exponential to get a damping rate. Consistently with this interpretation, we would expect the decay of $E_r \equiv -\phi'(r)$ to be determined by the return of neoclassical currents to an ambipolar equilibrium only delayed by the ion inertia through the ion polarization current. To make this more explicit we start from the momentum balance equation summed over species
\begin{equation}
	m_i\frac{\partial}{\partial t}n\bu_i + \nabla\cdot\Pi_i + \nabla\cdot\Pi_e + =  \bj\times\bB~,
\label{eq:mbe}
\end{equation}
where we keep only the ion's inertia ($m_e\ll m_i$). Projecting this equation on $\epol$ and taking flux surface averages we get
\begin{equation}
m_i\frac{\partial}{\partial t}\fsa{\epol\cdot n\bu_i} + \fsa{\epol\cdot\nabla\cdot\Pi_i} + \fsa{\epol\cdot\nabla\cdot\Pi_e} = \rev{\frac{\fsa{\bj\cdot\nabla\psi}}{2\pi}}=0~,
\label{eq:evol}
\end{equation}
where the quasi-neutrality condition $\nabla\cdot\bj = 0$ has been used to get the last identity. The first term on the LHS of Eq.\ref{eq:evol} is the ion polarization current, while the second and third terms are recognized as the ion and electron neoclassical non-ambipolar currents.

In the above discussion electron viscosity is usually neglected, for $\Pi_e/\Pi_i\sim \sqrt{m_e/m_i} \ll 1$ for equal ion and electron temperatures. While this approximation may be correct in ion-root plasmas, we need to keep electron viscosity and neoclassical flux to model our low density electron-root plasmas. In this parameter region, the ion and electron neoclassical fluxes display a similarly \emph{strong} $E_r$ dependence around its equilibrium value \cite{VelascoPPCF2012} and therefore both contribute to the return to the neoclassical ambipolarity.

The projection of the neoclassical pressure tensor is approximated in terms of the flow components:
\begin{equation}
\fsa{\epol\cdot\nabla\cdot\Pi_s} = \mu_p^sE_s + \mu_b^s\Lambda_s~,
\label{eq:visclinear}
\end{equation}
where $\mu_p^s$ and $\mu_b^s$ are the poloidal and bootstrap viscosities of respectively. We refer the reader to \cite{WobigPPCF2000} for further details. Analytical expressions for the viscosities coefficients in the plateau regime (the relevant regime for the edge of TJ-II plasmas) can be found in Ref.\cite{ShaingPoF1983}. These expressions were found to be in good agreement with the numerical solution of the drift-kinetic equation with DKES in LHD. A similar agreement has been found in TJ-II \cite{VelascoEFTC2011}. In this reference it is also shown that the poloidal viscosity term in Eq.\ref{eq:visclinear} is dominant in the edge of TJ-II. Using this approximation in Eq.\ref{eq:evol} we get
\begin{equation}
	m_in\fsa{\epol\cdot\epol}\frac{\partial E_i}{\partial t} = -(\mu_p^i +\mu_p^e) (E_i -  E_{i_0}) + R~.
\label{eq:evol2}
\end{equation}
Here,  \rev{$E_{i0}=\frac{2\pi}{ne}\frac{\mu_p^e}{\mu_p^e+\mu_p^i}(p_i'+p_e')$} represents the ambipolar ion poloidal flow obtained by zeroing the time-derivative of Eq.\ref{eq:evol}. $R$ represents other possible momentum fluxes (i.e. Reynolds' stress) not captured in Eq.\ref{eq:visclinear}. Therefore, our estimate of neoclassical damping rate is given by 
\begin{equation}
\nu_p = \frac{\mu_p^i+\mu_p^e}{m_i n\fsa{\epol\cdot\epol}}~.
\label{eq:nup}
\end{equation}
A usual approximation of the inertia term $\fsa{\epol\cdot\epol}$ \rev{for large aspect-ratio  and circular cross-section is} $\fsa{\epol\cdot\epol}\approx r^2(1+2q^2)$. In the case of TJ-II this approximation is only correct within a factor of 2. 
\begin{figure}
\includegraphics{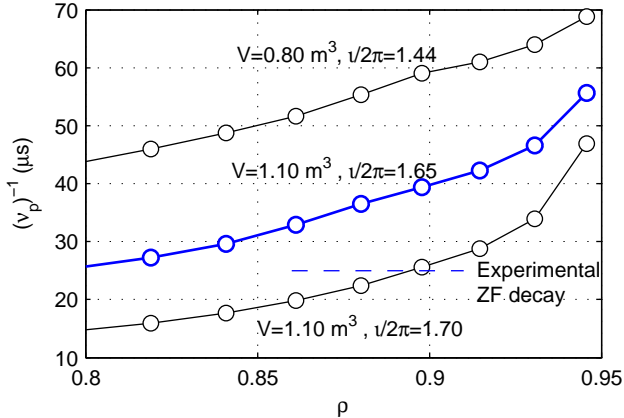}
\caption{Comparison of NC decay times with the observed ZF relaxation time. Calculations for three magnetic configurations (labeled with their volume and rotational transform at the last closed flux surface) are shown. The actual configuration used for the shots in fig.\ref{fig:profiles} is in blue. The radial decay of $\nu_p$ is mainly due to the decaying $T$-profiles. For the for these calculations we took $T_i(0.9) = 50$~eV and $T_e(0.9) = 120$~eV.}
\label{fig:nc}
\end{figure}
The poloidal viscosity is related to the non-ambipolar radial diffusion and was computed with DKES~\cite{HirshmanDKES}: the mono-energetic neoclassical coefficients were calculated following~\cite{SugamaPoP2002} and then convolved with a Maxwellian distribution describing the equilibrium plasma. Indeed, it can be seen that Eq.\ref{eq:visclinear} are equal to the equations proposed by Sugama and Nishimura \cite{SugamaPoP2002} for poloidal mass flow balance neglecting the parallel and poloidal heat fluxes (i.e. keeping only the particle flux terms). More precise calculations are underway. 
Also note that the calculated viscosities correspond to a fixed value of $E_r$, namely the ambipolar electric field calculated self-consistently \cite{VelascoPPCF2011}. We therefore assume that the viscosity nonlinearity is not too strong, as is to be expected in the low poloidal Mach number conditions discussed here. \rev{Finally, we must again note that the relatively fast ($\sim 10\mu$s) time-scales involved in the rise and decay of the potential structures would require a time-dependent treatment of the viscosity (i.e. without neglecting the explicit time derivative of the distribution function in the drift-kinetic equation \cite{ShaingPoP2005, SugamaPRL2005}).}

\rev{The comparison of the estimated NC decay times with the experimental values is shown in Figure \ref{fig:nc}. Despite the aforementioned approximations, a reasonable (within a factor~$\times 2$) agreement is found. A more systematic comparison would require local temperature and density measurements (to  be used as an input for the viscosity calculations) and the extension of these studies to magnetic configurations with different damping characteristics, like those shown in Fig.\ref{fig:nc}.} This is left to future work.

\subsection{Density profile modulations and comparison with density compression by the $E\times B$ flow}

\begin{figure}
\includegraphics{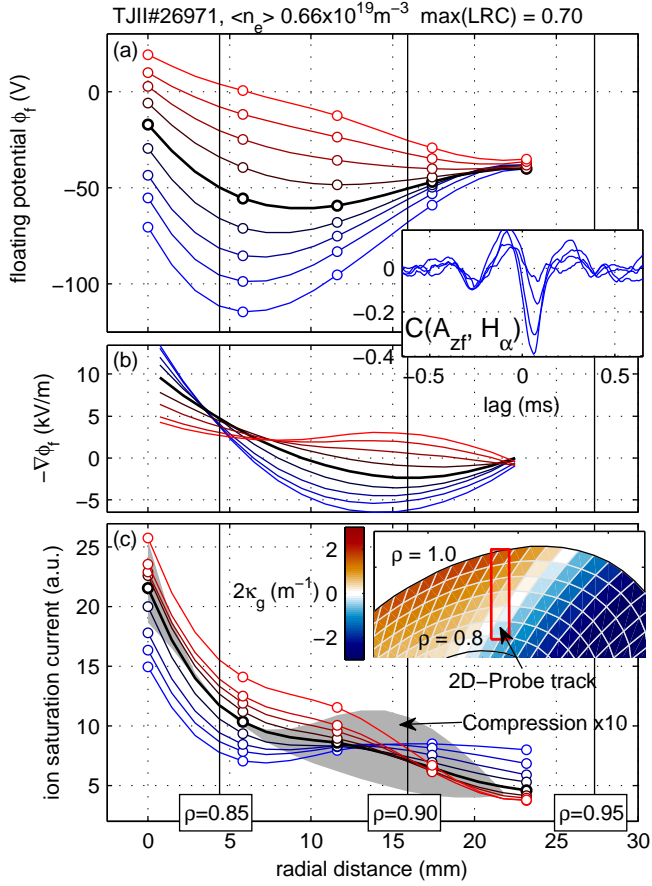}
\caption{Floating potential (a), approximated radial electric field (b) and Ion saturation current (c) modulations associated to the LRCed ZF-like structure in the edge of TJ-II. The inset in (a) shows the anti-correlation between the $H_\alpha$ radiation and $A_{ZF}$ first reported in \cite{AlonsoNF2012}. The inset in (c) shows the values of twice the geodesic curvature in the probe region. These were used to compute the expected $E\times B$ compressional density modulations shown in gray.}
\label{fig:isat}
\end{figure}

In Ref.\cite{AlonsoNF2012} we reported on the correlation between the ZF amplitude and the $H_\alpha$ monitors as a proxy of the outward flux of particles. Positive ZF amplitude (corresponding to a positive increment of $E_r$) was seen to be dynamically correlated with decreased levels of $H_\alpha$ radiation. This correlation is shown in figure \ref{fig:isat}.a (inset) for the shot no 26971. In this shot the 2D probe was operated with columns 1, 2 and 4 (see figure \ref{fig:probe}) operated in floating potential mode and column 3 set to ion saturation current ($I_{sat}$) mode. The extraction method outlined above was applied to the floating pins together with the distant probe potential to obtain the $\phi_f$ modulations shown in Fig.\ref{fig:isat}.a with estimated radial electric field $(-\frac{d}{dr}\phi_f)$ and $I_{sat}$ modulations shown in Fig.\ref{fig:isat}.b and c. 
The line averaged density is somewhat higher than for the shots shown in figure \ref{fig:profiles} and the average floating potential profile displays \rev{more negative values approaching the electron to ion root transition}. 

Ion saturation current depends mainly on plasma density and electron temperature $I_{sat}\propto n\sqrt{T_e}$, so its modulation is to be interpreted as modulations of these quantities with a dominant contribution from the density. We observe a steepening (flattening) of the $I_{sat}$ profile for positive (negative) values of $A_{ZF}$. This is consistent with the observed anti-correlation $C(A_{ZF}, H_\alpha)$. However we note that, in this shot, the floating potential profiles observed for negative $A_{ZF}$ amplitudes, exhibit large $E\times B$ shearing rates around $\rho= 0.85$, when and where the density profile flattens. 

As shown in figure \ref{fig:waveforms}.a, the ZF rise times are in the range of a few microseconds. Pressure waves transit time is $qR/c_s\sim 10 \mu s$ so it is appropriate to ask whether the observed modulations in density can be due to compressional effects. Compressional pressure sidebands and their associated parallel pressure gradients have been considered a mechanism of energy transfer (so-callled `geodesic transfer') from the zonal flow back to turbulence \cite{ScottNJP2005, NaulinPoP2005}. To first approximation the density sidebands $n_1(\psi, \theta,\phi)$ are expected to be given by the compression of the background flux-constant density $n_0(\psi)$, i.e.
\[
\frac{\partial n_1}{\partial t} \approx -n_0\nabla\cdot\mathbf{v_E}^{ZF} - \mathbf{v_E}\cdot\nabla n_1~,
\]
with the second term in the RHS being a factor $n_1/n_0$ smaller than the first. Compression of $\mathbf{v}_E = \frac{\bB\times\nabla\phi(\psi)}{B^2}$ is due to toroidicity $\nabla\cdot\mathbf{v}_E \approx -2\mathbf{v}_E\cdot\nabla\ln B\approx -2\mathbf{v}_E\cdot\bk$, where $\bk$ is the field line curvature and the approximations are good for low-$\beta$ plasmas. Note that the compression of the mean $E\times B$ velocity is compensated by the Pfirsch-Schlueter parallel flow, so we only consider the compression of the ZF velocity field.

The inset in figure \ref{fig:isat}.c shows the values of $2\kappa_g\equiv 2\frac{\nabla\psi\times\bB}{B|\nabla\psi|}\cdot\bk$, which according to the above discussion relate to the increments of density by
\begin{equation}\label{eq:comp}
\frac{\Delta n_1}{n_0} \approx 2\kappa_g\frac{E_r^{ZF}}{B}\Delta t~,
\end{equation}
where $E_r^{ZF} = -\frac{d\phi^{ZF}}{dr}$. The values of $2\kappa_g$ are positive in the outer probe region $\sim 0.5$ m$^{-1}$ go to zero around $\rho=0.85$ and turn slightly negative further inside. Figure \ref{fig:isat}.c shows the comparison of the actual density modulation with the $E\times B$ compression estimate given by Eq.\ref{eq:comp} with $\Delta t = 20\mu s$ (see Fig.\ref{fig:dynamics}) shown in gray. This estimate casts considerably smaller relative modulations (a few percent) compared to the observed $\sim 25\%$. It is well known (see \cite{WinsorPoF1968}) that the amplitude of density perturbations in geodesic acoustic modes are much smaller than their electric potential counterpart. Another prominent disagreement in this comparison is that the location of maximum $E_r$ modulation and expected density compression coincides with the radial position where $I_{sat}$ hardly varies. 

All the above observations indicate that the observed density modulations are not due to the compression of the background density by the $E\times B$ flow, but rather to radial transport of particles in agreement with our previous findings \cite{AlonsoNF2012}.

\section{Summary and Conclusions}
In this work we have studied the dynamics of the zonal-flow like structures responsible of the long range floating potential correlation observed in the TJ-II stellarator under certain conditions \cite{HidalgoEPL2009}. We made use of a previously introduced \cite{AlonsoNF2012} ZF extraction method to discuss possible drive (Reynolds stress) and damping (Neoclassical viscosity, geodesic transfer) mechanisms for the associated $E\times B$ velocity. We show that: (a) while the observed turbulence-driven forces can provide the necessary perpendicular acceleration, a causal relation could not be firmly established, possibly because of the locality of the Reynolds stress measurements. (b) The calculated neoclassical viscosity and damping times are comparable to the observed zonal flow relaxation times. The configuration dependence of this decay will be addressed in future work. (c) Although an accompanying density modulation is observed to be associated to the zonal flow, it is not consistent with the excitation of pressure sidebands, like those present in geodesic acoustic oscillations, caused by the compression of the $E\times B$ flow field.
 
 \section*{Acknowledgments}	
 J.A.A. wishes to acknowledge useful discussions with I.~Calvo,  V.~Naulin, E. S\'anchez, S.~Satake, H.~Sugama and H. Takahashi. The work of J.A.A. was funded by an EFDA Fusion Researcher Fellowship (Contract Nr. FU07-CT-2007-00050). 
 
\section*{References}
\providecommand{\newblock}{}

\end{document}